\documentclass[twocolumn,aps,showpacs]{revtex4}

\usepackage[dvips]{graphicx}
\usepackage{dcolumn}
\begin{document}
\title{Zipf's law in human heartbeat dynamics}
\author{J.\ Kalda, M.\ S\"akki}  
\affiliation{
Institute of Cybernetics,
Tallinn Technical University,
Akadeemia tee 21,
12618 Tallinn,
Estonia}
\author{M.\ Vainu}
\affiliation{Tallinn Diagnostical Center, P\"arnu mnt.\ 104, Estonia}
\author{M.\ Laan}
\affiliation{N\~omme Children Hospital, Laste 1, Tallinn, Estonia}

\begin {abstract} 

It is shown that the distribution of low variability periods in the activity of
human heart rate typically follows a multi-scaling Zipf's law. The presence or 
failure of a power law, as well as the values of the scaling exponents, are 
personal characteristics depending on the daily habits of the subjects.
Meanwhile, the distribution function of the low-variability periods as a whole 
discriminates efficiently between various heart pathologies. This new technique
is also applicable to other non-linear time-series and reflects these aspects
of the underlying intermittent dynamics, which are not covered by other 
methods of linear- and nonlinear analysis.

\end {abstract} 

\maketitle

The nonlinear and scale-invariant aspects of the heart rate
variability (HRV) have been studied intensively during the last
decades. This continuous interest to the HRV can be attributed to the
controversial state of affairs: on the one hand, the nonlinear
and scale-invariant analysis of HRV has resulted in many methods of
very high prognostic performance (at least on test groups) \cite{Amaral,Ivanov,Thurner,Saermark};
on the other hand, practical medicine is still confident to the 
traditional ``linear'' methods.
The situation is quite different from what has been observed three decades ago,
when the ``linear'' measures of HRV became widely used as important
noninvasive diagnostic and prognostic tools,
soon after the pioneering paper \cite{Hon}.
Apparently, there is a need for further evidences for the superiority of
new methods and for the resolution of the existing ambiguities.

During recent years the main attention of studies has been
focused on the analysis of the scale-invariant methods. It has been
argued that measures related to a certain time-scale (e.g.\ 5 min) are
less reliable, because the characteristic time-scales of
physiological processes are patient-specific. The scale-invariant
measures are often believed to be more universal and sensitive to life-threatening 
pathologies \cite{Amaral,Ivanov}. However, carefully designed time-scale-related 
measures can be also highly successful, because certain
physiological processes are related to a specific time scale \cite{Thurner}. 

The scale invariance has been exclusively seen in the heart rhythm
following the (multi)fractional Brownian motion (fBm) \cite{Peng}. It
has been understood that the heart rhythm fluctuates in a very
complex manner and reflects the activities of the subject (sleeping,
watching TV, walking etc.) \cite{Poon,Struzik} and cannot be adequately described by
a single Hurst exponent of a simple fBm. In order to reflect
the complex behavior of the heart rhythm, the multi-affine
generalization of the fBm has been invoked \cite{Amaral,Ivanov}; it has been claimed that 
the multifractal scaling exponents are of a significant prognostic value.

The approach based on fBm
addresses long-time dynamics of the heart rhythm while completely neglecting 
the short-scale dynamics on time scales less than one minute (the respective 
frequencies are typically filtered out \cite{Peng}). The {\em short-time variability} 
has been described only by the so called linear measures, such as 
$p_{NN50}$ (the probability that two adjacent normal heart beat intervals 
differ more than 50 milliseconds).
Meanwhile, the level of the short-time variability of the human heart rate varies
in a very complex manner, the high- and low-variability periods are deeply intertwined \cite{Poon}. 
This 
is a very important aspect, because the low-variability periods are the 
periods when  the heart is in a stressed state, with high level of signals arriving 
from the autonomous nervous system.
The conventional linear measures are not appropriate for describing
such a complex behavior. Thus, there is a clear need for suitable nonlinear methods.

In this paper we present a new scale-invariant description of the 
short-time variability of the heart rate.
It is shown that the distribution of low-variability
periods in the activity of a normal heart follows the Zipf's law. 
It is also shown that the distribution function of the low-variability
periods contains a considerable amount of diagnostically 
valuable information. This new 
technique is also applicable to other non-linear 
time-series, such as EEG signals and financial data \cite{KSK}.

Our analysis is based on ambulatory Holter-monitoring data 
(recorded at Tallinn Diagnostic Centre) of 218 patients with various diagnoses.
The groups of patients are shown in Table 1. The sampling rate of ECG was 
180~Hz. 
The patients were monitored  during 24 hour under normal daily activities.
The preliminary analysis of the ECG recordings was performed using the 
commercial software; this resulted in the sequence of the {\em normal-to-normal} (NN)
intervals $t_{NN}$ (measured in milliseconds), 
which are defined as the intervals between two subsequent normal heartbeats (i.e.\ normal QRS complexes).
\begin {table}[!bth]
\begin {tabular}{|l|c|c|c|c|c|c|c|}
\hline
         & Healthy & IHD & SND  & VES & PCI & RR & FSK\\
 \hline
No.\ of 
patients & 103    & 8   & 11   & 16  & 7   & 11 & 6 \\
\hline
Mean age & $45.5$ & $65.4$ & $50.0$ & $55.9$ & $47.3$ & $55.5$ & $11.7$\\
\hline
Std.\ dev.\ 
of age & $20.5$ &  $11.4$ & $19.3$ &  $14.3$ &  $11.6$ &  $14.4$ &  $4.6$ \\
\hline
\end{tabular}
\caption{
Test groups of patients. Abbreviations are as follows: IHD - Ischemic Heart Disease (Stenocardia); SND - Sinus Node Disease; 
VES - Ventricular Extrasystole; PCI - Post Cardiac Infarction; RR - Blood Pressure Disease; 
FSK - Functional Disease of Sinus Node.}
\end{table}

Originally, the Zipf's law addressed the distribution of words in a language \cite{Zipf}:
every word has assigned a rank, according to its  ``size'' $f$, 
defined as the relative number of occurrences in some long text 
(the most frequent word obtains rank $r=1$, the second frequent --- $r=2$, etc).
The empirical size-rank distribution law $f(r) \propto r^{-\alpha}$
is surprisingly universal:
in addition to all the tested natural languages, it applies
also to many other phenomena.  
The scaling exponent $\alpha$ is often close to one 
(e.g.\ for the distribution of words).
Typically, the Zipf's law is applicable to a
dynamical system at statistical equilibrium, when the following conditions are
satisfied: {\em (a)} the system consists of elements of different size; {\em (b)}
the element size has upper and lower bounds; {\em (c)} there is no intermediate 
intrinsic size for the elements.
As already mentioned, the human heart rhythm has a complex structure, where 
the duration $\tau$ of the low-variability periods varies in a wide 
range of scales, from few to several hundreds of heart beats.
Thus, one can expect that the 
distribution of the low-variability periods follows the Zipf's law
\begin {equation}
r \propto \tau ^{-\gamma}.
\end {equation}
However, the scaling behavior should not be expected to be perfect.
Indeed, the heart rate is a non-stationary signal affected by the non-reproducible daily 
activities of the subjects. The non-stationary pattern of these activities, together with their time-scales,
is directly reflected in the above mentioned distribution law.
This distribution law can also have a fingerprint of the characteristic time-scale (around ten to twenty seconds) 
of the blood pressure oscillations. 
Finally, there is a generic reason why the Zipf's law fails (or is non-perfect) at small rank numbers.
The Zipf's law is a statistical law; meanwhile, each rank-length curve is 
based on a single measurement. Particularly, there is only one longest low-variability period (and likewise, only one 
most-frequent word), the length of which is just as long as it happens to be, there is no averaging whatsoever.
For large rank values $r$, the relative statistical uncertainty can be estimated as $1/\sqrt{r}$.

To begin with, we 
define the local variability for each ($i$-th) interbeat interval 
as the deviation of the heart rate from the local average, 
\begin {equation}
\delta(i) = \frac {|t_{NN}(i)-\left<t_{NN}(i) \right>|}{\left<t_{NN}(i) \right>}.
\end {equation}
The angular braces denote the local average, 
calculated using a narrow (5 beats wide) Gaussian weight function. Further, we introduce a 
threshold value $\delta_0$; $i$-th interbeat interval is said to have a low variability,
if the condition
\begin {equation}
\delta(i) \le \delta_0
\end {equation}
is satisfied.
A {\em low-variability period}  is defined as a set of consecutive 
low-variability intervals; its length $\tau$ is measured in the number of
heartbeats. Finally, all the low-variability periods are arranged according to their lengths and
associated with ranks. 
The rank of a period is plotted versus
its length in a logarithmic graph, see Fig.~1;
Zipf's law would correspond to a straight descending line.

\begin {figure}[!bth]
\includegraphics{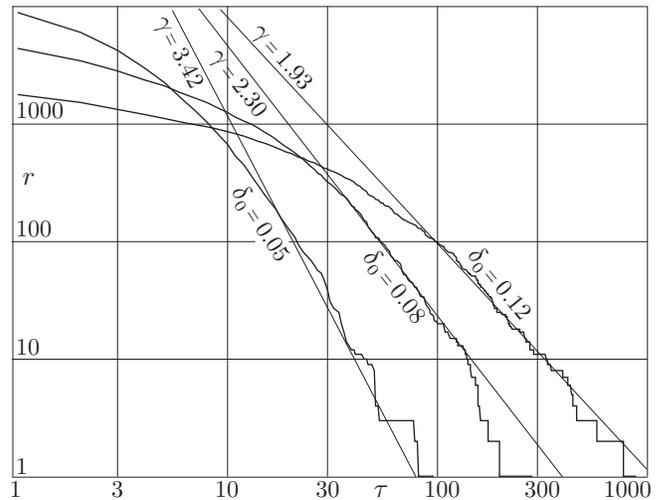}
\caption{Multi-scaling distribution of the low-variability periods: the rank $r$ of a period is 
plotted versus its duration $\tau$ (measured in heartbeats) for different values of the threshold parameter 
$\delta_0$.}
\end{figure}
For a very low threshold parameter $\delta_0$, 
all the low-variability periods are very short, because it is difficult to satisfy the stringent condition (3).
In that case, the inertial range of scales is too short for a meaningful scaling law. On the other hand,
for a very high value of $\delta_0$, there is a single low-variability period occupying the entire
HRV-recording. Between these two 
cases, there is such a range of the values of $\delta_0$, which 
leads to a 
non-trivial rank-length law.
For a typical healthy patient, the $r(\tau)$-curve 
is reasonably close to a straight line, and the scaling exponent $\gamma$ is  a function
of the threshold parameter $\delta_0$. Thus, unlike all the other well-known applications of the Zipf's law, 
we are dealing with a {\em multi-scaling law}. 

Recently, Ivanov et.\ al.\ \cite{Ivanov} have reported that anomalous multifractal spectra
of the HRV signal indicate an increased risk of sudden cardiac death. 
Therefore, it is natural to ask, does the presence or failure 
of the multiscaling behavior indicate the healthiness of the patient?
In what follows we discuss a somewhat more general question: what is the relationship between 
the properties of the distribution function of the low variability periods and the diagnosis 
of the patient. Testing the prognostic significance 
for predicting sudden cardiac death, which is also of a great importance, has been
postponed due to the nature of our test groups.

First, let us analyze the correlation between the diagnosis of a patient and the 
scaling exponent $\gamma$.
To begin with, we have to determine the optimal value for the threshold parameter $\delta_0$. 
For a meaningful analysis, the 
scaling behavior should be as good as possible.
It turned out that for a typical patient, the best approximation of the 
function $r(\tau)$ with a power law is achieved for $\delta_0 \approx 0.05$ (see Fig.\ 2a); 
in what follows, all the values of the exponent $\gamma$ are calculated for $\delta_0 = 0.05$.
It should be noted that for some
patients, the length-rank distribution is still far from a power law (see Fig.\ 2b).
\begin {figure}[!b]
\includegraphics{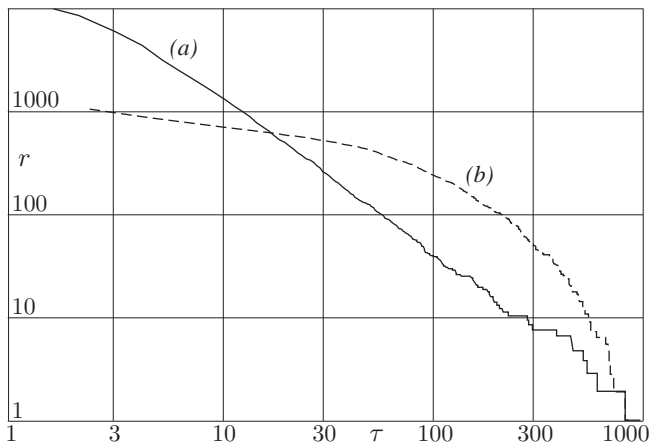}
\caption{Rank-length curves for a patient with a good power law {\em (a)} and for a patient 
with no power law {\em (b)}. In both cases, the threshold parameter $\delta_0=0.05$.}
\end{figure}

The slope of a curve on the logarithmic plot is calculated using root-mean-square (rms)
fit for such a range of lengths 
$[\tau_{\mbox{\scriptsize start}}, \tau_{\mbox{\scriptsize end}}]$, for which 
the $r(\tau)$-curve is nearly a power law, and the scaling range width
$\Delta=\ln \tau_{\mbox{\scriptsize end}} - \ln \tau_{\mbox{\scriptsize start}}$
is as large as possible.
Bearing in mind the statistical nature of the Zipf's law and non-stationarity of
the underlying signal, we have chosen a not 
very stringent definition of  what is ``nearly a power law'', see Fig.~3.
Around the rms-fit-line, two limit lines are drawn; 
$\tau_{\mbox{\scriptsize start}}$ and $\tau_{\mbox{\scriptsize end}}$ correspond 
to the points, where the $r(\tau)$-curve crosses the limit lines.

\begin {figure}[!b]
\includegraphics{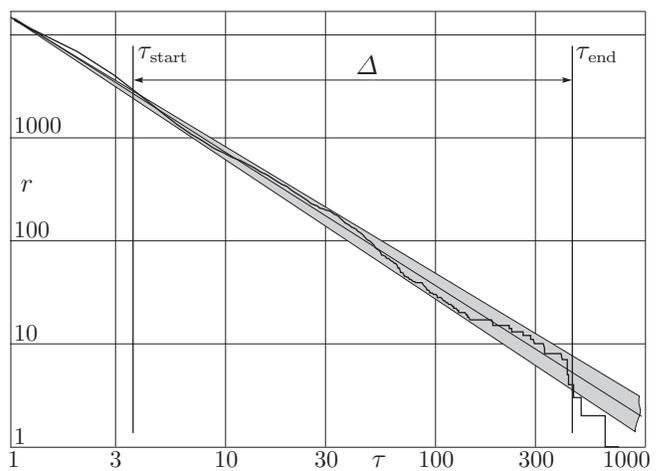}
\caption{Definition of the width of the scaling interval $\Delta$. The rank-length curve is fitted with 
a power law; the boundaries of the scaling interval are defined as the 
intersection points of limit lines and $r(\tau)$-curve.}
\end{figure}
Note that the precise placement and shape of the limit lines is arbitrary,
i.e.\ small variations do not lead to qualitative effects.
Here, the distance of the limit lines from the central line has been chosen to be 
$\ln 2$ at $\tau = \tau_{\max}$, and zero at $\tau = 1$,
where $\tau_{\max}$ is the length of the longest low-variability period. 
Admitting mismatch  $\ln 2$ at $\tau = \tau_{\max}$ is motivated by the observation that 
due to the lack of any statistics, 
the longest low-variability period could have been easily twice as long as we measured it to be.
For large rank values, the statistical uncertainty is assessed as $\sqrt{r}$;
in logarithmic graph, this would correspond to an exponentially decreasing (with increasing $r$) 
distance between a limit curve and the central fit-line. However, the above mentioned
effect of the non-stationary pattern of the subjects daily activities makes the situation 
more complicated. There is no easy way to quantify this effect and therefore, we opted for the simplest 
possible solution, simple straight limit lines.

The scaling exponent $\gamma$ has been calculated for all the patients and 
Student test was applied to every pair of groups. In most cases, the 
significance was quite low; two best distinguishable groups were RR and FSK, the
result of Student test being 5.7\%. Therefore, one can argue that 
the slopes of linear parts are highly personal characteristics depending also on the
daily habits of the subjects, which are weakly correlated with diagnosis.

Further we tested, how is the failure of the power law correlated with the diagnosis.
The width of the scaling range $\Delta$
was 
used as a measure of how well the curve is corresponding to a power law. 
The Student test results for 
the parameter $\Delta$ turned out to be  similar
to what has been observed for the parameter $\gamma$: the correlation between the 
failure of the power law and diagnosis was weak. Thus, a rank-length curve
resembling the one depicted by a dashed line in Fig.~2, does not hint to
heart pathology. 
It should be also noted that the dashed curve in Fig.~2 can be considered as a {\em generalized 
form of scale-invariance} with scale-dependent  differential scaling exponent.
Such a 
behavior 
seems to be universal; for instance, certain forest fire models \cite{Chen} 
lead to the differential fractal dimension depending on the local scale.

Finally, we analyzed the diagnostic significance of the parameters $\ln \tau_{\mbox{\scriptsize end}}$  
and $\ln \tau_{\mbox{\scriptsize start}}$. 
This analysis does make sense, because typically, 
the start- and end-points of the scaling range correspond to 
certain physiological time-scales. 
The parameter $\ln \tau_{\mbox{\scriptsize end}}$
provided, indeed, a remarkable resolution between the groups of patients, see Table 2.
According to the Student test, the
healthy patients, were distinct from five heart pathology groups with  probability
$p<1.6$\%. The parameter $\ln \tau_{\mbox{\scriptsize start}}$
was diagnostically less significant.
\begin{table}[!tb]
\includegraphics{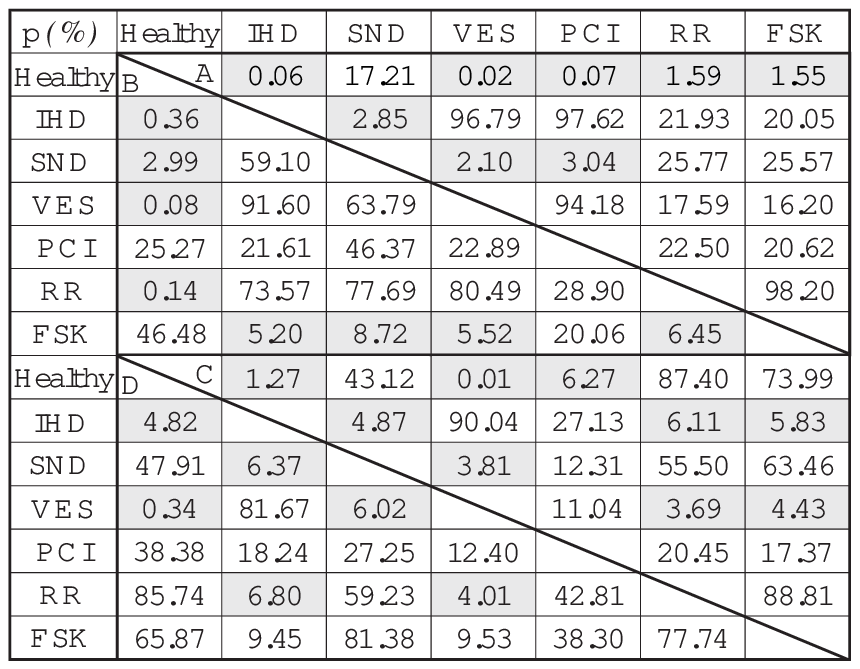}
\caption{$p$-values of the Student test.
Data in the topmost triangular region (with label {\em A}) are calculated using 
the parameter $\ln \tau_{\mbox{\scriptsize end}}$. Triangular region {\em B} corresponds to the parameter 
$\ln r_{\max}$, region {\em C} --- to $\ln r_{100}$, and region
{\em D} --- to  $\ln \tau_{40}$. Gray background highlights small $p$-values, $p < 10\%$.}
\end{table}
\begin{figure}[!!tb]
\includegraphics{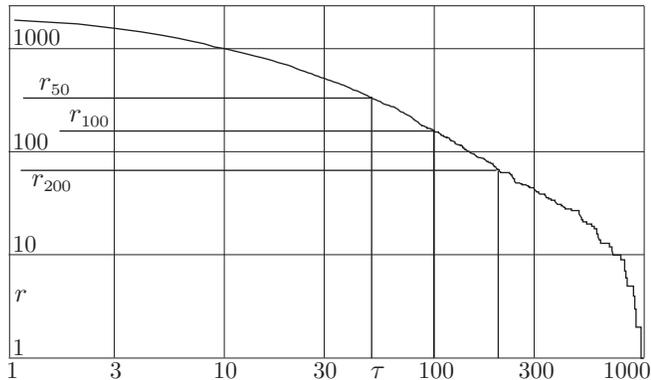} 
\caption{Definition of the parameters $r_{50}$, $r_{100}$, and $r_{200}$.}
\end{figure}

Unfortunately, the calculation of the parameter $\tau_{\mbox{\scriptsize end}}$ is technically quite 
a complicated task, not suited for clinical practice. Therefore, we aimed to 
find a simpler alternative to it.
Basically, the strategy was to find a simple parameter reflecting the behavior of
the rightmost (large-$\tau$) part of the $r(\tau)$-curve.
An easy option is $\ln \tau_{\max}$, which has been already analyzed \cite{JKMS}.
This parameter has indeed a considerable diagnostic value, but
its reliability is decreased by the above discussed statistical fluctuations.
Better alternatives are provided by {\em (a)} the overall number of low-variability periods  $r_{\max}$ 
(which is small, if there are lot of long low-variability periods); 
{\em (b)} the coordinates of specific points of the rank-length curve.
Here  we chose a set of critical ranks $R =10$, $20$ or $40$, and determined the respective
lengths $\tau_R$ so that $r(\tau_R)=R$. We also fixed  a set of critical 
length values, $T=50$, $100$, or $200$, and determined the respective
rank numbers $r_T = r(T)$, see Fig.\ 4.
Both techniques
turned out to be of high diagnostic performance;
illustrative $p$-values are given in Table 2. 
Parameters $\tau_{10}$ and $\tau_{20}$ 
performed less well than  $\tau_{40}$ (for instance, 
the $p$-values for the healthy and VES-subject groups were 0.60\%, 0.58\% and 0.34\%, respectively),
and are not presented in tabular data.
Similarly, $r_{100}$ turned out to be more efficient than $r_{50}$ and $r_{200}$
(the respective healthy and VES-group $p$-values  being $0.02$\%, $0.01$\%, and $0.09$\%).
It also outperforms $\tau_{40}$, but is sometimes less efficient than $r_{\max}$ 
or $\tau_{\mbox{\scriptsize end}}$ (see Table 2).
Hence, various heart pathologies seem to affect the 
heart rate dynamics at the time scale around 100 heart beats (one to two minutes).

In conclusion, new aspect of non-linear time-series has been discovered,
the scale-invariance of low-variability periods.
We have shown that the distribution of low variability periods in the activity of
human heart rate typically follows a {\em multi-scaling Zipf's law}. 
The presence or failure of a power law, as well as 
the values of the scaling exponents, are personal characteristics depending 
on the daily habits of the subjects.
Meanwhile, the distribution function of the low-variability periods as a whole 
contains also a significant amount of diagnostically valuable information, the most part of 
which is reflected by the parameters $r_{100}$, $r_{\max}$, and $\tau_{\mbox{\scriptsize end}}$, see
Table 2. 
These quantities characterize the complex structure of 
HRV signal, where the low- and high variability periods are deeply intertwined, aspect which 
is not covered by the other methods of heart rate variability analysis (such as 
fractional Brownian motion based multifractal analysis). As a future development, it would be of 
great importance to analyze the prognostic value of the above mentioned parameters for patients with sudden cardiac death.

The support of ESF 
grant No.\ 4151 is acknowledged.

\end{document}